# Title: Airborne Biomarker Localization Engine (ABLE) for Open Air Point-of-Care Detection

**Authors:** Jingcheng Ma[1,2*], Megan Laune[3], Pengju Li[4,5], Jing Lu[6,7], Jiping Yue[3], Yueyue Yu[6,7], Jessica Cleary[8], Kaitlyn Oliphant[6,7], Zachary Kessler[9], Erika C. Claud[6,7*], Bozhi Tian[1,3,5*]

**Affiliations:**
[1]The James Franck Institute, University of Chicago, Chicago, IL, 60637.
[2]Department of Aerospace and Mechanical Engineering, University of Notre Dame, Notre Dame, IN, 46556.
[3]Department of Chemistry, University of Chicago, Chicago, IL, 60637.
[4]Pritzker School of Molecular Engineering, University of Chicago, Chicago, IL, 60637.
[5]The Institute for Biophysical Dynamics, University of Chicago, Chicago, IL 60637.
[6]University of Chicago Medical Center, Chicago, IL, 60637.
[7]Comer Children's Hospital, University of Chicago, Chicago, IL, 60637.
[8]Duchossois Family Institute, University of Chicago, Chicago, IL, 60637.
[9]Former address: The College, University of Chicago, Chicago, IL, 60637.

*Corresponding author. Email: jma9@nd.edu, eclaud@bsd.uchicago.edu, btian@uchicago.edu.

**Abstract:** Unlike biomarkers in biofluids, airborne biomarkers are dilute and difficult to trace. Detecting diverse airborne biomarkers with sufficient sensitivity typically relies on bulky and expensive equipment like mass spectrometers that remain inaccessible to the general population. Here, we introduce Airborne Biomarker Localization Engine (ABLE), a simple, affordable, and portable platform that can detect both volatile, non-volatile, molecular, and particulate biomarkers in about 15 minutes. ABLE significantly improves gas detection limits by converting dilute gases into droplets by water condensation, producing concentrated aqueous samples that are easy to be tested. Fundamental studies of multiphase condensation revealed unexpected stability in condensate-trapped biomarkers, making ABLE a reliable, accessible, and high-performance system for open-air-based biosensing applications such as non-contact infant healthcare, pathogen detection in public space, and food safety.

**Main Text**

Portable biosensors allow onsite detection of biomarkers for disease management from just a few droplets of biofluids like blood and sweat(*1-5*). Airborne biomarkers also have important implications in public health(*6*), non-invasive disease diagnosis(*7-11*), and food safety(*12*), but sensors for airborne biomarkers are far less developed. Simultaneous detection of diverse airborne biomarkers with sufficient sensitivity heavily relies on bulky and expensive equipment like the mass spectrometers, limiting the wide usage of gas biosensing to the general population.

Current techniques for detecting airborne biomarkers face two major challenges: sensitivity and versatility. For molecular biomarkers, many can be as dilute as ppb or ppt level in the open air(*8*), but miniature gas sensors often have much lower sensitivity (~ 1 ppm, **Supplementary Table 1**)(*13*). Recent innovations integrate miniature gas sensors on wearable device for convenient personal usage(*11, 14*), but the intrinsic sensitivity limit remains to be significantly improved. For particulate biomarkers (*e.g.*, bacteria and pollens), simple devices like filters and impactors cannot capture particles below a certain hydrodynamic diameter(*15*). "Condensation Growth Tube (CGT)" method cool down sample air within a narrow tube, initiating steam condensation on environmental particles to increase their size, allowing the capture of particles as small as 5 nm(*16-18*). However, such homogenous nucleation design requires expensive hardware (~$10,000) to draw in and process sample air in a closed system. In addition, a general platform that simultaneously capture and detect diverse biomarkers is often needed for general populations, but neither current molecular nor particulate biosensors can achieve such flexibility.

**Molecular Biomarker Transformation through Multicomponent Condensation**

To address these two challenges, we report a universal platform, Airborne Biomarker Localization Engine (**ABLE**), that can trap a wide range biomarkers from dilute air into concentrated droplets through dropwise condensation on nanostructured condensers (**Fig. 1A**). Such heterogeneous nucleation can be realized by a small condensing chamber (5-10 cm in size), much cheaper and simpler than mass spectrometers. ABLE enables 7-10 orders of magnitudes enhancement in the detection limit by allowing the detection of airborne biomarkers using well-developed liquid biosensors from high performance lab equipment to widely accessible paper-based test strips.

For molecular biomarkers, liquid biosensors are far more advanced than gas biosensors and can sometimes detect analytes at pM level (**Fig. 1B**). A noticeable characteristic of fluids biosensors is the low sample volume requirement (10 µL)(*19, 20*). If airborne biomarkers are concentrated from air (volume $V_a \approx 1$ m³ and low concentration $C_a \approx 1$ ppb) into a single droplet (volume $V_l \approx 10$ µL), the resultant droplet would have a high concentration ($C_l$):

$$C_l \approx 10^{-9} \frac{V_a C_a}{V_l V_M} = 4.5 \text{ mM}. \qquad (1)$$

Where $V_M$ is the molar volume of air. As many commercially available test strips can achieve a detection limit of 1 µM for physiologically relevant biomarkers like glucose and lactate acid(*4, 21*), 4.5 mM is considered very concentrated. It becomes obvious that instead of sensing dry gas, utilizing gas-liquid conversion and use liquid biosensors to detect airborne molecules can be a promising alternative biosensing mechanism (**Fig. 1B**).

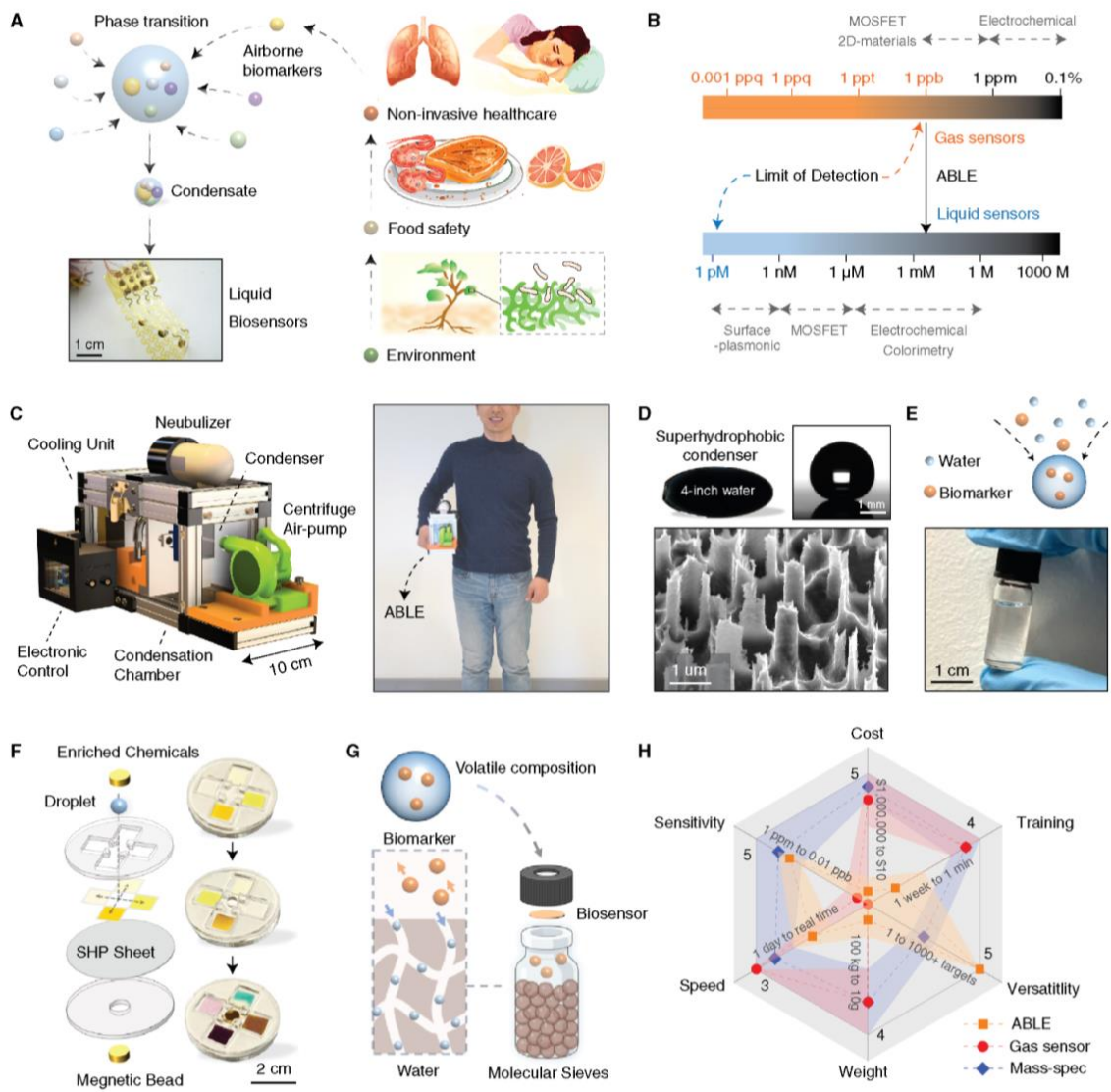

**Fig. 1: Concept, design, and device components of Airborne Biomarker Localization Engine (ABLE).** **(A)** Schematic diagram of ABLE operation. Dilute airborne biomarkers are transformed into concentrated droplets and can be detected using readily available liquid sensors. **(B)** Comparative analysis of the detection limit between ABLE and existing gas sensors for dilute molecular biomarkers in the air. MOSFET stands for Metal-Oxide-Semiconductor Field-Effect Transistor. **(C)** Portable ABLE device design, with condensation chamber size 8.8 cm × 10.8 cm × 8.4 cm and adjacent water-cooling chamber, DC power control box, and air-pump. **(D)** Optical image, contact angle, and surface morphology of the superhydrophobic condenser materials, which consists of black silicon with layer of HMDS and $CF_x$ fabricated on silicon wafer. **(E)** Schematic diagram of multiphase condensation and photo of ~2 mL vial of condensate sample after a 15-minute air sampling. **(F)** Schematic diagram and optical image of portable readout element with colorimetric array for detection of non-volatile chemicals in enriched droplets. **(G)** Schematic diagram of the 3Å molecular sieve used to enrich and detect volatile chemicals. **(H)** Comparative analysis of performance and cost-effectiveness of gas sensing between portable gas sensors, mass spectroscopy, and ABLE.

To achieve such biomarker gas-liquid conversion, we exploit a ubiquitous but underexplored physical phenomenon: multicomponent vapor condensation. Atmospheric air consists mostly of nitrogen, oxygen, and ~3% water vapor when saturated. When atmospheric moisture nucleates into liquid condensates upon flowing across a cold surface, assuming all biomarkers are simultaneously trapped inside, there exists a linear relationship, $C_l(\mu M) \approx 3C_a(ppb)$, between the biomarker's concentration in the liquid condensate ($C_l$) and in the air ($C_a$) (**Supplementary Text 1**). So the hypothetical gas-to-liquid conversion described in **Equation 1** may be experimentally realized by a condensation process (from ppb to μM) and optional liquid enrichment process (from μM to mM). For multicomponent condensation, $N_2$ and $O_2$ in the air are known to be "non-condensable" and will be ejected from the condensate by convection and diffusion(*22, 23*). Here, we explore if "condensable" airborne impurities will also be trapped in the droplets, and how the thermodynamic efficiency varies between molecules of different polarity and volatility.

**Hardware Design of Airborne Biomarker Localization Engine**

In **Fig. 1C**, we outline the hardware design of ABLE, which contains a ~10 cm condensing chamber, a miniature air pump, a nebulizer, and a plane condenser cooled by a Peltier cooler at 3°C ± 2 °C (**Supplementary Text 2, Supplementary Fig. S1, S2**), and can be made under $200 (**Supplementary Table 3**). We adopted a thermally conductive superhydrophobic surface design for the condenser, a nanostructured black silicon substrate coated with an ultra-thin (<100 nm-thick) water repellent coating consists of a hexamethyldisilazane (HMDS) monolayer and fluorinated ($CF_x$) polymer layer (**Fig. 1D, Methods, Supplementary Fig. S3**). The superhydrophobic surface allows "jumping-droplet condensation", which allows a 250% condensation compared to filmwise condensation (**Fig. 2A-2B, Supplementary Text 3-4, Supplementary Fig. S2**)(*24, 25*). The superhydrophobic surface also ensures minimized adhesion with chemical residues. The prototype developed for demonstration has an airflow rate of 18 L/min and can collect around 1 mL condensate in 10 minutes (**Fig. 1E**). ABLE with different sizes could also be made using similar designs, depending on the specific applications.

Once airborne biomarkers are trapped in the condensates, further enhancement can be achieved by water reduction. For non-volatile biomarkers, water evaporation allows sample enrichment from dilute solution into concentrated residue, which can be rehydrated into concentrated solution (**Fig. 1F, Supplementary Text 5**). For volatile chemicals, we developed a molecular sieve-based enrichment setup. The micro-vial contains 1 g of 3Å molecular sieve beads and colorimetric test strips on the underside of the vial cap (**Fig. 1G, Methods**). When the dilute solution is added to the vial, water is adsorbed into the 3Å molecular sieve, and the biomarkers are released in airborne state and encapsulated in a 0.2 mL head space. With such design strategy, ABLE demonstrates superior performance (**Fig. 1H;** detailed grading definitions in **Methods**) and can reliably detect both molecular and particle biomarkers with flexibility.

**Experimental Determination of Thermodynamic Efficiency ($\eta$)**

The most important metric for ABLE is thermodynamic efficiency ($\eta$), defined as the ratio between measured condensate concentration $C_{l,cond}$ to the theoretical maximum. $\eta = 1$ represents a biomarker-to-water molecular ratio the same as that in gas. Quantitative determination of $\eta$ allows precise $C_a$ measurement through liquid sensors that measures $C_{l,cond}$:

$$C_{l,cond}(\mu M) = 2.48 \frac{\eta RT}{p_v} C_a (\text{ppb}). \tag{2}$$

Where $R$ is gas constant, $T$ is temperature, and $p_v$ is saturated vapor pressure.

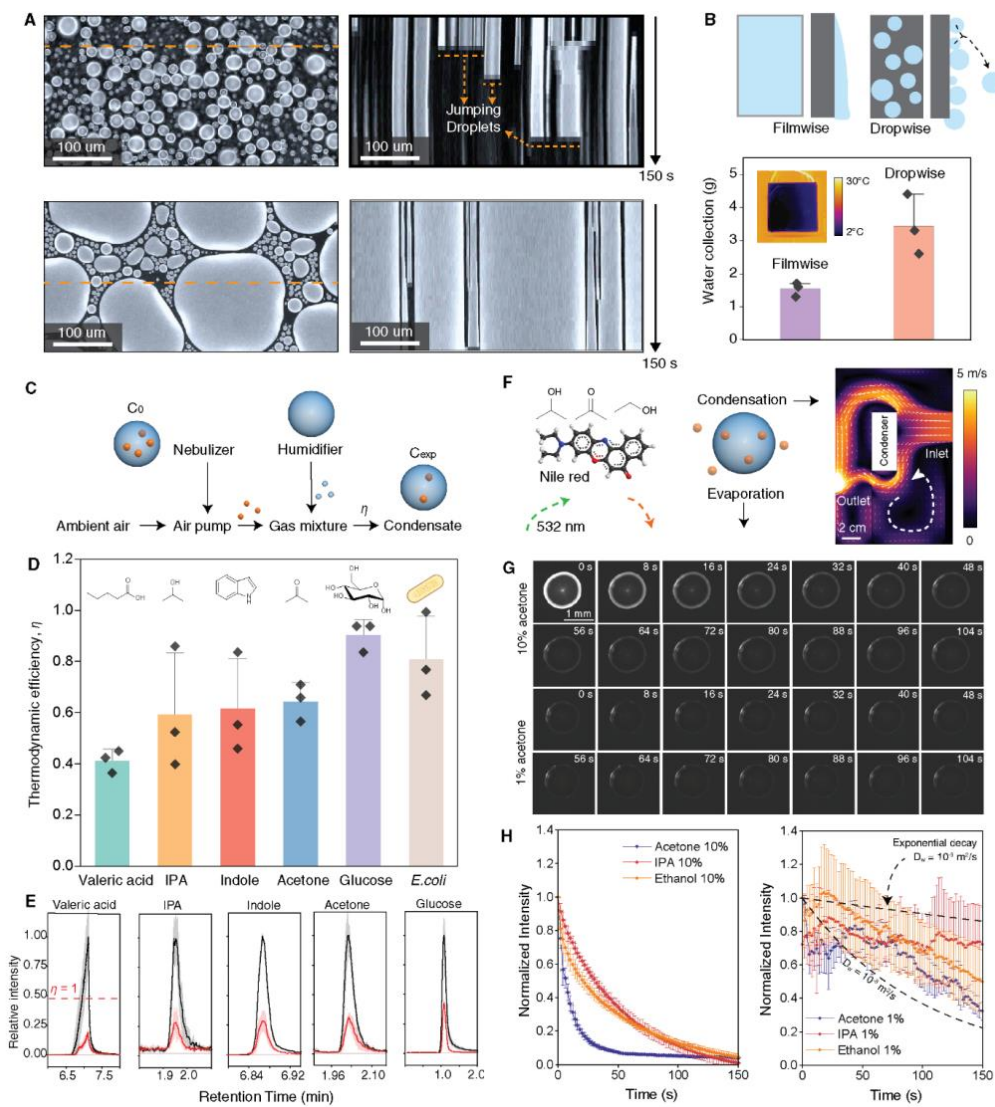

**Fig. 2: Thermofluidic design principles of ABLE.** **(A)** Microscopic optical image and kymograph of water condensation on nanostructured black silicon (top; with jumping-droplet condensation) and polished silicon (bottom). **(B)** Schematic diagram of dropwise condensation, filmwise condensation, and experimentally determined water collection rates from a 30-minute collection ($N = 3$ independent measurements). Error bars extend to the outliers. Inset: temperature profile of the ambient and the 4 cm × 4 cm condenser surface. **(C)** Experiment design for determining thermodynamic efficiency for controlled chemicals. **(D)** Thermodynamic efficiency of ABLE for various chemicals. Data are expressed as mean ± s.d. of $N = 3$ independent samples. **(E)** Mass spectroscopy measurements of chemical abundance in benchmark

experiments. Black curves represent measurements from the standard aqueous solution. Red curves represent measurements following application of ABLE. Mean value ± s.d. of *N* = 3 independent samples are indicated by the shaded areas. **(F)** COMSOL simulation and experimental investigation of sample stability during condensation and evaporation phases. **(G)** Representative fluorescence intensity of 10% (v/v) and 1% (v/v) binary acetone-water droplets evaporating under ambient conditions. **(H)** Evaporation kinetics of binary droplets at different concentrations. Mean value ± s.d. of *N* = 5 independent samples.

We studied how molecular biomarker's polarity and volatility affects $\eta$. A standard system is developed to experimentally determine $\eta$ using a combination of aerosolization process and mass spectroscopy (**Fig. 2C, Methods, Supplementary Fig. S4-S5**). We describe biomarkers' volatility using Henry's constant in water. The polarity of chemicals is described by their water solubility. Standard gases including acetone (Henry's constant $k_H = 1.6 \times 10^{-2}$ mol/m³·Pa in water(*26*)), isopropyl alcohol ($k_H = 5.4 \times 10^{-2}$ mol/m³·Pa(*26*)), valeric acid ($k_H = 8.6 \times 10^{-1}$ mol/m³·Pa(*26*)), indole ($k_H = 5.4 \times 10^{-1}$ mol/m³·Pa(*26*)), and glucose (non-volatile) at a concentration of 14 ppm. These chemicals were produced in the airborne state *via* aerosolization from standard liquid solutions, and then pumped into ABLE as standard gas. With a Stokes number $St \sim O(10^{-3})$, the flow condition is designed so that all biomarkers should closely follow the inlet air stream (**Methods**). The ratio between the concentration of standard liquid solution ($C_0$) and condensate ($C_{exp}$) are determined by mass spectrometer with a theoretical expectation of:

$$\frac{C_{exp}}{C_0} = \frac{4RTU_{neub}}{p_{H_2O}\pi D^2 U_f \rho_{M, H_2O}} \eta. \tag{3}$$

Where $U_{neub}$ represents the mass flux of the nebulizer, $U_f$ is airflow velocity, $D$ is the exit diameter of the air pump, and $\rho_{M, H_2O} = 18$ g/mol. In our experimental system, $C_{exp}/C_0 = 0.47\eta$ (**Methods**), allowing direct measurement of $\eta$.

We do not consider gas solubility to be a limiting factor of $\eta$ because most gas are dissolvable in water when dilute (<10 ppb) (**Supplementary Fig. S6**). We did expect that at equilibrium, Henry's law presents a theoretical limit in $\eta$ (0.01 – 1, **Supplementary Text 9, Supplementary Fig. S7**). However, as shown in **Fig. 2D-2E**, we find that $\eta$ is surprisingly high and consistent across all standard gases tested. For non-volatile and highly water-soluble glucose, $\eta \sim 1$, which indicates effective capture. Although the consistency in $\eta$ is promising for biosensing performance, the consistent $\eta \sim 0.5$ across all volatile chemicals with large differences in Henry's constant was unexpected and requires further investigation.

**Stability and Evaporation Kinetics of Multicomponent Condensates**

Loss of volatile chemicals may occur either during condensation when airflow is on, or during sample collection stage when airflow is stopped and volatiles in the condensate start to evaporate. As shown in **Fig. 2F**, our numerical simulation of airflow suggests a vortex flow that returns evaporated chemicals to the inlet air stream for re-condensation (**Methods, Supplementary Fig. S8**), so we do not expect a significant loss in $\eta$ at this stage. We therefore contribute the observed consistency in $\eta \sim 0.5$ to a consistent evaporation rate. However, this assumption opposes the commonly held belief that the evaporation rate of a volatile phase from water binary droplet is strongly associated with the chemical's Henry's constant in water(*27-29*).

We hypothesize that the unexpected $\eta$ only happens when 1) the evaporation timescale is sufficiently short, otherwise Henry's law would dictate $\eta \sim 0$ for volatiles; and 2) the analyte concentration is sufficiently low, otherwise both Henry's constant and solubility would affect evaporation kinetics (which we did not observe). We therefore experimentally test if the low analyte concentration contributes to the stable and consistent evaporation pattern.

Utilizing water binary droplets with acetone, IPA, as well as ethanol (Henry's constant $k_H = 1.1 \times 10^{-1}$ mol/m$^3$·Pa in water(*26*)) as the standard system, we observe the evaporation physics using Solvatochromism Fluorescence Microscopy (**Methods, Supplementary Fig. S9**). Fluorescence images of droplets with as high as 10% volume concentration show a chemical-dependent evaporation, consistent with the literature (**Fig. 2G-2H**)(*30*). As shown in **Fig. 2G**, acetone is concentrated at the solid-liquid interface, indicating outward Marangoni flow that accelerates near-surface evaporation(*28*). However, in the dilute regime where chemical concentration < 5%, the evaporation of acetone, ethanol, and isopropyl alcohol becomes much slower and largely overlaps, as if Marangoni flow no longer dominates the evaporation kinetics (**Fig. 2G-2H, Supplementary Fig. S10-S11**). It is noticeable that the relative concentration $\bar{C} = 0.5 \pm 0.1$ after around one minute of evaporation, which indeed corresponds to $\eta \sim 0.5$ and cross verifies the consistent thermodynamic efficiency of ABLE.

It is known that Marangoni flow may be inhibited if surface tension gradient falls below a certain level(*28*). We hypothesize that evaporation of dilute binary droplets, and perhaps dilute multicomponent condensates in general, shows a distinct evaporation physics wherein mass diffusion becomes the solo governing parameter. Since the diffusion coefficient of different molecules in water spans a very narrow range $D_w \sim 10^{-9}$ to $10^{-8}$ m$^2$/s (*31*), it is expected that these chemicals in droplets evaporate on a similar timescale $\tau_D \sim L^2/D_w$. Given that $L \sim 10^{-3}$ m in our experimental system (**Supplementary Fig. S11**), $\tau_D \sim 10^2$ to $10^3$ seconds. In **Fig. 2H**, we utilize a first-order approximation of $I \sim e^{-t/\tau_D}$ to describe the evaporation kinetics, which indeed corresponds well with the experimental data and supports our hypothesis. The unexpected evaporation physics highlights the robustness of ABLE for detection of a wide range of molecular biomarkers. The stability of different chemicals is related almost only to their mass diffusion coefficient in water; therefore, the description of ABLE efficiency is dramatically simplified when working with volatile biomarkers that may have very different physicochemical properties (**Supplementary Fig. S12**).

**Sample Enrichment Strategies and Performances for Dilute Targets**

Airborne biomarkers from human exhaled condensates (EBC) contains many physiologically relevant metabolites(*7*), but many are very dilute and requires sample enrichment for downstream sensors. The compatibility of the condensate sample with colorimetric test strips is especially important for ABLE usage by the general population. We therefore conducted a benchmark experiment where artificial EBCs with controllable chemical composition and concentration was made to test the response of commercially available test strips (**Fig. 3A** and **Supplementary Table 2**). 10 μL of artificial EBCs were drop casted onto the test strips and the corresponding color change (ΔE) was sequentially measured (**Fig. 3B**). We found that 1× EBC produced an ΔE < 3 that is imperceptible to most people. But 10× and 100× enrichment of artificial EBCs produce readable color signals for most chemicals (**Fig. 3C-3D**), which thus becomes the goal of enrichment.

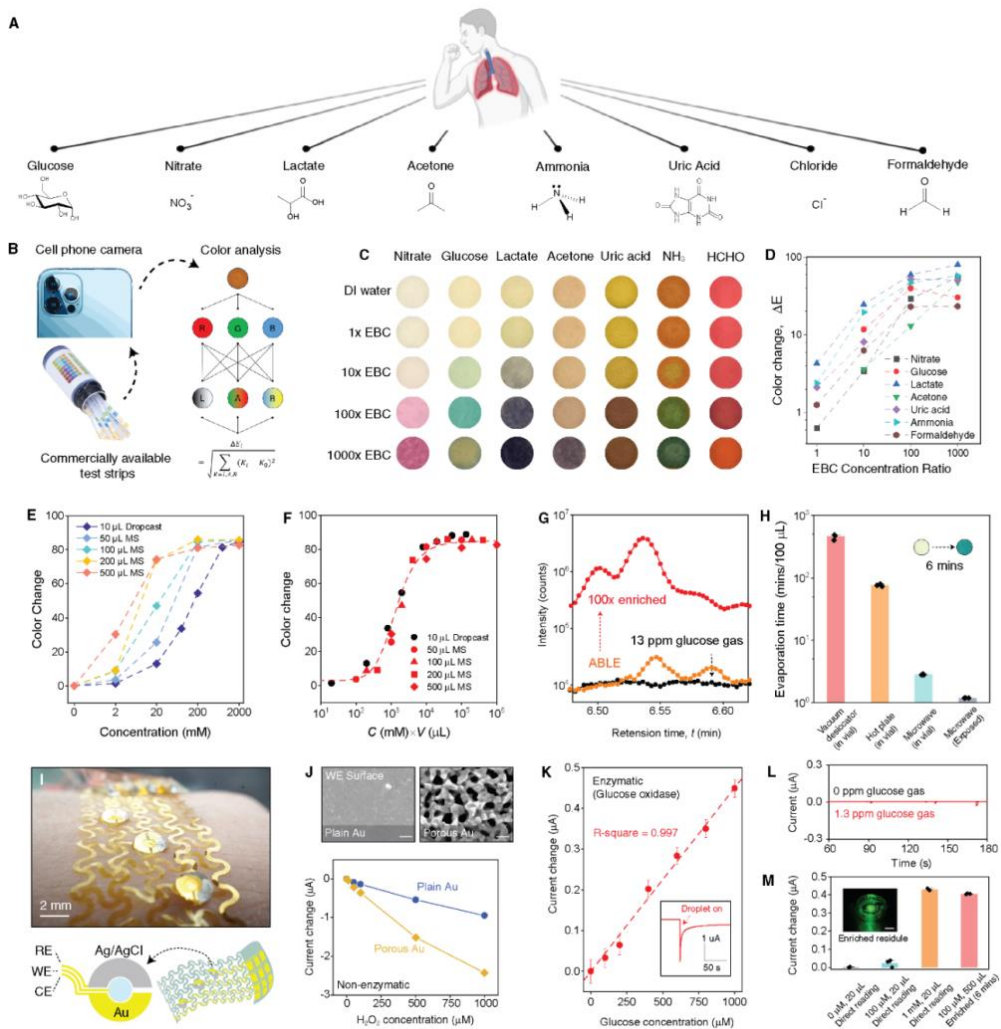

**Fig. 3: Enrichment strategies for detecting dilute biomarkers.** **(A)** Schematic diagram of exhaled metabolites and the chemical composition of artificial EBCs used in benchmark experiments. **(B)** Sample processing workflow. Airborne metabolites are captured into droplets, which are detected by colorimetric test strips. Color change is quantified in CIE-LAB color space to determine chemical abundance. **(C)** Optical image of color strips following addition of 10 µL artificial EBCs at different concentrations ($N = 1$). **(D)** Quantified color change of test strips in (C) when compared to DI water groups ($N = 1$). **(E-F)** Color change of ketone test strips using molecular sieve enrichment with different sample volume compared to drop-cast method (10 µL sample), including both **(E)** raw data and **(F)** super-positioned data. Standard acetone solutions were used for benchmark. **(G)** Mass spectroscopy data showing the intensity of airborne glucose absorbed on empty sample tube (black), tube with 1 µL of ABLE collected glucose condensate before (orange) and after (red) enrichment, and **(H)** time required to enrich non-volatile samples for hot plate evaporating, microwave evaporating, and vacuuming methods. Insert: enrichment of 10 mg/L glucose water solution using microwave evaporating. **(I)** Photograph and schematic diagram of flexible electrochemical sensors. **(J)** Design and fabrication of plain gold and porous gold working electrode for sensitivity enhancement. Uncertainty for each point is determined by the standard deviation of $N = 3$ measurements, and are too small to be shown. SEM scale bar: 100 nm. **(K)** The benchmark characterization

of enzymatic glucose sensors. Uncertainty for each point is determined by the standard deviation of $N = 3$ measurements. **(L-M)** enhanced performance when coupled with ABLE, including **(L)** enabling the reading of airborne chemicals, without ABLE there would be no signal with or without glucose presented in the dry air, and **(M)** amplified signals through sample enrichment treatment.

For volatile samples, the molecular sieve method demonstrated reliable readout enhancement with standard acetone solutions, showing immediate (~ 1 minute) 5×, 10×, 20×, and 50× of enrichment effects **(Fig. 3E-3F)**. For non-volatile biomarkers, direct water evaporation is effective. We tried different methods including direct heating on hot plate, water evaporation under vacuum, and microwave heating. The use of microwave oven appears to be the fastest and most home-accessible method (~100 µL/minute, **Fig. 3G-3H, Supplementary Fig. S17**)(*32*).

The gas-to-liquid conversion and further enrichment process can also convert electrochemical liquid sensor into gas sensors. As a demonstration, we developed representative nano-electrochemical sensors with a design where nano-porous gold, created through microfabrication and a combination of micelle-templated metal deposition(*33*) on both rigid and flexible substrates (**Supplementary Text 10, Supplementary Fig. S18-S21**). Porosity can be introduced onto the working electrode to increase the sensitivity of the bioelectronics (**Fig. 3I-3J**). The enzymatic glucose sensor is further fabricated by enzyme deposition, with a linear detection range from 0.1 mM to 1 mM (**Fig. 3K**) when using a standard PBS glucose solution. As such sensors require an aqueous electrolyte environment, no readable current can be generated from a standard glucose gas, highlighting the need of ABLE for such sensing purpose (**Fig. 3L**). ABLE also facilities sensitive readings due to its capability of sample enrichment. For dilute samples, such as a 0.1 mM glucose solution that barely produces a signal for our system, 6-minutes of microwave enrichment from a 500 µL solution results in dry glucose residue that can be rehydrated, showing significant current enhancement (**Fig. 3M**).

**The Detection of Dilute Biomarkers in Open Air**

ABLE can perform open-air biosensing for non-contact disease management. Such capability is important because it serves vulnerable populations when physical contact between the medical device and patients is either not feasible or should be limited. In addition, it can monitor public health at a population level, complementing current individual detecting methods. Compared to technologies such as wearable breath sensors that create a closed system(*11*), open-air sensing requires additional consideration on dilution factors that concerns the open-space dilution, the mismatch between the water content and air flux in the ABLE chamber and the exhalation source. A linear correction model is proposed here that can connect the apparent biomarker concentration in ABLE condensates and the ones in EBC (**Fig. 4A**, **Supplementary Text 11** and **Supplementary Fig. S22**). The reliability of the model is demonstrated by human experiments that detect ppt-level glucose from human exhaled breath and yields an expected Blood:EBC glucose concentration ratio of $10^4$:1(*8, 34*) (**Supplementary Text 12-13, Supplementary Fig. S23-S24**).

We highlighted ABLE's non-contact healthcare capability for preterm infants, a vulnerable population that are born before completing 37 weeks of gestation and acquire severe morbidities in when comparing to term infants(*35, 36*). Early-stage disease diagnosis for preterm infants have been challenging and non-invasive methods are desired(*37*). Often, preterm infants are housed in isolette chambers with circulated air. The chemical composition of this air is underexplored as a potential route to obtain biomarkers from lung or gastrointestinal track for disease diagnosis.

To navigate this possibility, we conducted animal experiments with "humanized" mouse model (**Fig. 4B**, **Methods**). We orally gavaged microbiome communities from human preterm infants into germ-free mouse dams as gut microbiome is strongly associated with the level of underdevelopment(*38-40*). This process resulted in offspring that acquired a microbiome from the dam that represented the original human infant's microbiome(*38*). Two clinically relevant microbial communities are used: 1) Preterm: microbial communities from preterm infants born at 24 weeks of gestational age with risks for lung and gut deficits (N = 5 mouse, individual weight of 17.2 g ± 1.0 g); and 2) Term: microbial communities from full term healthy infants at 39 weeks of gestational age (N = 7 mouse, individual weight of 14.8 g ± 0.9 g). At four weeks of age, the offspring were put in a cage (Innocage® mouse cage) with ABLE for 45 minutes of air collection. The volume ratio between the mouse cage (~7 L) and an isolate (~130 L) is 1:19, similar to the weight ratio between the mouse group (~ 100g) and a preterm infant (~ 2000g).

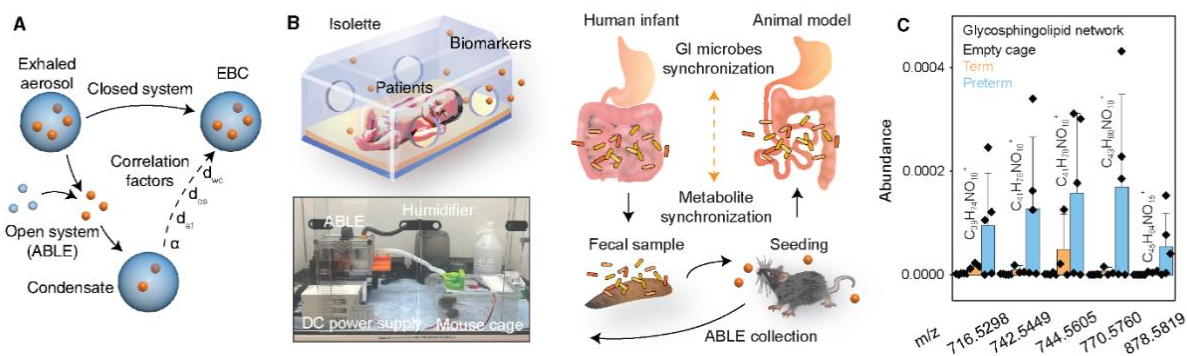

**Fig. 4: Open air detection of disease biomarkers. (A)** Schematic diagram of detecting exhaled biomarkers in open and closed system. Dilution factors need to be considered when using ABLE in an open system. **(B)** Schematic diagram, animal model, and experimental setup of open-air biomarker capture for preterm infants' disease diagnosis. **(C)** The relative abundance of features that were putatively identified as glycosphingolipids in the sample air of empty mouse cage ($N = 5$), cage of preterm mouse ($N = 5$) and term mouse ($N = 3$). Data are expressed as mean ± s.d.

To identify the very dilute biomarkers in the condensates, we performed untargeted mass spectrometry-based metabolomics and analyzed the data using molecular networking and *in silico* structure predictions (**Supplementary Text 14**). Multiple samples (1 mL each) were collected from the same batch of preterm and term mouse and sent for metabolite analysis. Various biomolecules, such as glycosphingolipids, were readily detected and found to be more abundant in the preterm samples (**Fig. 4C**). Sphingolipids are emerging as integral modulators of airway inflammation, immune cell infiltration, and airway hyperresponsiveness(*41, 42*), which are common symptoms seen in preterm infants(*43*). Recent literature using asthma animal model has shown that glycosphingolipids in the lung were upregulated in the asthma groups and the levels of glycosphingolipids were positively significantly associated with airway inflammation and eosinophilia(*44*). The glycosphingolipids found in the ABLE samples shown here, although dilute, suggest that sphingolipids can also be found in the open air for non-invasive detection. This study can serve as the starting point of systematically analysis of other potential airborne metabolite and help diagnosing challenging preterm diseases using chemical information of air in open space.

**Detection of Mesoscopic Particle Biomarkers**

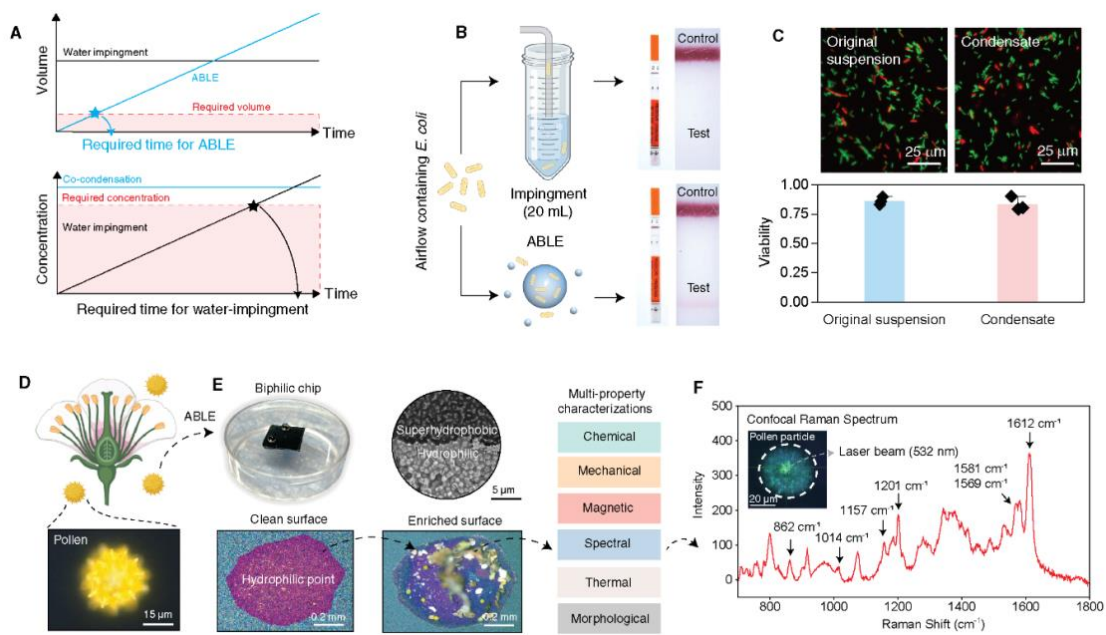

**Fig. 5: Detection of particle biomarkers.** **(A)** The principle of ABLE operation allows faster detection of airborne microbes than traditional water impingement methods. **(B)** ABLE demonstrates high *E. coli* detection sensitivity compared with the state-of-the-art water impingement method. Both methods used commercially available lateral flow assays to detect the presence of airborne *E. coli* at a density of ~$10^4$ CFU/m$^3$. **(C)** ABLE maintains high bacterial viability due to its gentle and rapid collection process ($N = 3$ independent measurements at different views). Error bars extend to the outliers. **(D)** Schematic of capturing airborne particles for in-situ, multi-property characterizations when using pollen particles as example. **(E)** Sample condensates collected by ABLE is evaporation-enriched on biphilic surface, allowing the deposition of pollen and other particles from sample plant onto a ~1 mm spot. **(F)** Raman confocal characterization of the capture pollen with well-studied peaks outlined. Inset: microscopic image of the pollen particle illuminated by a laser spot (532 nm).

For many mesoscopic biomarkers (*e.g.* pollen, bacteria, or virus), lateral flow assays are widely used for personal diagnosis of infectious disease(*19*), but techniques for large-scale public health monitoring are underdeveloped. Many particle biosensors collect biomarkers on a collector with fixed volume, and it can take days of time for biomarkers to accumulate or culture to reach satisfying concentration. ABLE works completely differently by collecting samples at a constant concentration, and it is the sample volume increases over time. It is therefore possible to use ABLE for rapid collection of biomarkers that rapidly achieve satisfying concentration (**Fig. 5A**).

We used *E. coli* bacteria as a standard system to demonstrate the performance of ABLE for particle biomarker collection. We determined that $\eta$ was 70%-100% (**Supplementary Fig. S28**), similar to what was observed in non-volatile molecular biomarker glucose. Cell viability assays showed that *E. coli* maintained high viability before and after condensation, despite the potential oxygen exposure and osmotic shock(*45*) (**Fig. 5C**). We further compared the performance of ABLE against that of the conventional water impingement technique using standard $2.2 \times 10^4$ CFU/m$^3$ *E. coli* aerosol (representative of *E. coli* aerosol levels close to that found in hospital settings)(*46*). Traditional water impingement methods pump airborne microbes into 10-20 mL of water to create a dilute suspension, which is often not concentrated enough and requires further culturing process. In contrast, the bacteria condensate harvested from ABLE show higher concentration during a 20 minute collection interval, and it clearly shows a significant color change (line appearance) on the lateral flow test strip (**Fig. 5B,** QUICK$^{TM}$ 15 Minute Water Bacteria Test)(*47*).

ABLE's gas-liquid transformation capability also enables the deposition of airborne particles into concentrated residues on wafers, allows for various characterization methods. We demonstrated obtaining detailed chemical spectral information from airborne pollen allergens from Assorted Mums that are harvested ABLE (~3 cm away). Evaporation enrichment of the particle-containing condensate on a biphilic surface effectively concentrated the sample into a ~1 mm spot **(Fig. 5D-5E)**. The biphilic surface is made by manually depositing ~1 mm hydrophilic spot onto the condenser materials by Sharpe pan indenting. With such sample configuration, we can quantitatively understand spectral information of the airborne samples, such as using Raman confocal microscopy (well-characterized pollens peaks(*48, 49*) are shown in **Fig. 5F**). For complex and unknown components, recent work has shown the possibility of integrating multiple materials properties with machine-learning-based pattern recognition for biomarker identification (*21, 50-53*), and ABLE can enable unknown airborne biomarkers to be seen by these systems.

**Discussion**

ABLE presents a simple and cost-effective method for facile detection of a wide range of dilute airborne biomarkers, including both molecular and particle biomarkers. Leveraging multicomponent phase transition, this new paradigm for gas sensing allows development of ABLE variations with a wide range of choices in airflow rate, device size and mobility, and chemical phase selection for effective condensation collection. Based on ABLE's design principles, several future developments can enhance gas biosensing in terms of both portability and sensitivity. For faster sample collection, designing complex condensers for effective atmospheric moisture capture is crucial(*54, 55*), as the current vertical plane condenser design achieves only ~30% mass capture efficiency (**Supplementary Text 3**), highlighting a space for 300% improvement in sampling rate. In addition to water condensation, incorporating chemical compositions, such as surfactants for capturing non-polar molecules, may improve thermodynamic capture efficiency. Condenser surface coatings or hydrogel-based absorbent materials can also be optimized to selectively capture molecules with different physicochemical properties(*56-59*).

Although the applications shown here are mostly concerning healthcare, ABLE can also be used in many more applications such as food spoilage detection by the detection of biogenic amine (**Supplementary Text 15 and Supplementary Fig. S29, S30**). In addition, combining ABLE with more advanced liquid sensor designs, such as 2D materials(*60, 61*) or ultra-sensitive fluorescence probe(*62-64*) allows achieving unprecedented detection limit for gas biomarkers in the future.

# References and Notes

**Acknowledgments:** The authors thank K. M. Watters for scientific editing of the manuscript. The authors also thank Emile Augustine from the University of Chicago "Engineering + Technical Support Group" for the 3D modeling and fabrication of ABLE device. The authors also acknowledge that AI-assisted technology "ChatGPT 4o" was used to polish the writing and correct grammar mistakes during the preparation of the manuscript.

**Funding:**
US Army Research Office (W911NF-24-1-0053)
University of Chicago startup grant
University of Notre Dame startup grant
National Institutes of Health grant R01 HD105234
National Institutes of Health grant R21 NS121432
University of Chicago Grier Prize for Innovation Research in the Biophysical Sciences

**Author contributions:**
Conceptualization: JM, BT
Methodology: JM, ML, PL, JL, YY, JC, KO, ECC, BT
Investigation: JM, ML, PL, JL, JY, YY, JC, KO, ECC, BT
Funding acquisition: BT, ECC
Project administration: JM, BT, ECC
Supervision: BT, ECC
Writing – original draft: JM, ML, PL
Writing – review & editing: All authors

**Diversity, equity, ethics, and inclusion:** With the understanding that about half of the world's population has limited access to essential health services, this work aims to make efforts to ensure ABLE can be a technology that empower people to manage their health independent of location or socioeconomic status using the air around them.

**Competing interests:** A provisional patent application has been submitted through the University of Chicago. JM, PL, ZK, and BT are listed as the inventors. No competing interests have been declared by the others.

**Data and materials availability:** All data presented in this work are fully available from the corresponding authors upon reasonable request.

**Supplementary Materials:**
Materials and Methods
Supplementary Text 1-15
Figs. S1 to S30

Tables S1 to S3
References

# References

1. J. Wu, H. Liu, W. Chen, B. Ma, H. Ju, Device integration of electrochemical biosensors. *Nature Reviews Bioengineering* **1**, 346-360 (2023).
2. W. Y. Liu, H. Y. Cheng, X. F. Wang, Skin-interfaced colorimetric microfluidic devices for on-demand sweat analysis. *Npj Flexible Electronics* **7**, (2023).
3. Z. L. Ye *et al.*, A Breathable, Reusable, and Zero-Power Smart Face Mask for Wireless Cough and Mask-Wearing Monitoring. *Acs Nano* **16**, 5874-5884 (2022).
4. W. Gao *et al.*, Fully integrated wearable sensor arrays for multiplexed in situ perspiration analysis. *Nature* **529**, 509-+ (2016).
5. Y. Xiang, Y. Lu, Using personal glucose meters and functional DNA sensors to quantify a variety of analytical targets. *Nature Chemistry* **3**, 697-703 (2011).
6. J. V. Puthussery *et al.*, Real-time environmental surveillance of SARS-CoV-2 aerosols. *Nature Communications* **14**, (2023).
7. I. Horvath *et al.*, Exhaled breath condensate: methodological recommendations and unresolved questions. *European Respiratory Journal* **26**, 523-548 (2005).
8. D. Tankasala, J. C. Linnes, Noninvasive glucose detection in exhaled breath condensate. *Translational Research* **213**, 1-22 (2019).
9. C. Arnold, Diagnostics to take your breath away. *Nature Biotechnology* **40**, 990-993 (2022).
10. S. L. Zhang *et al.*, Rapid Measurement of Lactate in the Exhaled Breath Condensate: Biosensor Optimization and In-Human Proof of Concept. *Acs Sensors* **7**, 3809-3816 (2022).
11. W. Heng *et al.*, A smart mask for exhaled breath condensate harvesting and analysis. *Science* **385**, 954-961 (2024).
12. E. Istif *et al.*, Miniaturized wireless sensor enables real-time monitoring of food spoilage. *Nature Food* **4**, 427-436 (2023).
13. S. G. Chatterjee, S. Chatterjee, A. K. Ray, A. K. Chakraborty, Graphene-metal oxide nanohybrids for toxic gas sensor: A review. *Sensors and Actuators B-Chemical* **221**, 1170-1181 (2015).
14. T. Klotz, A. Ibrahim, G. Maddern, Y. Caplash, M. Wagstaff, Devices measuring transepidermal water loss: A systematic review of measurement properties. *Skin Research and Technology* **28**, 497-539 (2022).
15. G. Mainelis, Bioaerosol sampling: Classical approaches, advances, and perspectives. *Aerosol Science and Technology* **54**, 496-519 (2020).
16. S. V. Hering, M. R. Stolzenburg, A method for particle size amplification by water condensation in a laminar, thermally diffusive flow. *Aerosol Science and Technology* **39**, 428-436 (2005).
17. S. V. Hering, S. R. Spielman, G. S. Lewis, Moderated, Water-Based, Condensational Particle Growth in a Laminar Flow. *Aerosol Science and Technology* **48**, 401-408 (2014).
18. S. Asadi *et al.*, Influenza A virus is transmissible via aerosolized fomites. *Nat Commun* **11**, 4062 (2020).
19. R. Gupta *et al.*, Ultrasensitive lateral-flow assays via plasmonically active antibody-conjugated fluorescent nanoparticles. *Nature Biomedical Engineering*, (2023).
20. A. Koh *et al.*, A soft, wearable microfluidic device for the capture, storage, and colorimetric sensing of sweat. *Science Translational Medicine* **8**, (2016).
21. C. H. Xu *et al.*, A physicochemical-sensing electronic skin for stress response monitoring. *Nature Electronics*, (2024).
22. D. G. Kroger, W. M. Rohsenow, Condensation heat transfer in the presence of a non-condensable gas. *International Journal of Heat and Mass Transfer* **11**, 15-26 (1968).
23. R. S. Wang, J. H. Guo, E. A. Muckleroy, D. S. Antao, Robust silane self-assembled monolayer coatings on plasma-engineered copper surfaces promoting dropwise condensation. *International Journal of Heat and Mass Transfer* **194**, (2022).
24. J. Ma, S. Sett, H. Cha, X. Yan, N. Miljkovic, Recent developments, challenges, and pathways to stable dropwise condensation: A perspective. *Applied Physics Letters* **116**, (2020).

25. N. Miljkovic, R. Enright, E. N. Wang, Effect of Droplet Morphology on Growth Dynamics and Heat Transfer during Condensation on Superhydrophobic Nanostructured Surfaces. *ACS Nano* **6**, 1776-1785 (2012).
26. R. Sander, Compilation of Henry's law constants (version 4.0) for water as solvent. *Atmospheric Chemistry and Physics* **15**, 4399-4981 (2015).
27. A. A. Pahlavan, L. S. Yang, C. D. Bain, H. A. Stone, Evaporation of Binary-Mixture Liquid Droplets: The Formation of Picoliter Pancakelike Shapes. *Physical Review Letters* **127**, (2021).
28. J. R. E. Christy, Y. Hamamoto, K. Sefiane, Flow Transition within an Evaporating Binary Mixture Sessile Drop. *Physical Review Letters* **106**, (2011).
29. A. K. Bell *et al.*, Concentration gradients in evaporating binary droplets probed by spatially resolved Raman and NMR spectroscopy. *Proceedings of the National Academy of Sciences of the United States of America* **119**, (2022).
30. C. Diddens, Y. X. Li, D. Lohse, Competing Marangoni and Rayleigh convection in evaporating binary droplets. *Journal of Fluid Mechanics* **914**, (2021).
31. J. M. P. Q. Delgado, Molecular diffusion coefficients of organic compounds in water at different temperatures. *Journal of Phase Equilibria and Diffusion* **28**, 427-432 (2007).
32. P. Y. Zhao *et al.*, Multiphysics analysis for unusual heat convection in microwave heating liquid. *Aip Advances* **10**, (2020).
33. H. Lim *et al.*, A universal approach for the synthesis of mesoporous gold, palladium and platinum films for applications in electrocatalysis. *Nature Protocols* **15**, 2980-3008 (2020).
34. A. A. Karyakin *et al.*, Non-invasive monitoring of diabetes through analysis of the exhaled breath condensate (aerosol). *Electrochemistry Communications* **83**, 81-84 (2017).
35. S. Saigal, L. W. Doyle, An overview of mortality and sequelae of preterm birth from infancy to adulthood. *Lancet* **371**, 261-269 (2008).
36. L. L. Richter *et al.*, Temporal trends in neonatal mortality and morbidity following spontaneous and clinician-initiated preterm birth in Washington State, USA: a population-based study. *BMJ Open* **9**, e023004 (2019).
37. R. B. A. Butler, *Preterm Birth: Causes, Consequences, and Precention*. R. B. A. Butler, Ed., (National Academies Press, Washington, 2007), pp. 741.
38. L. Lu *et al.*, Transcriptional modulation of intestinal innate defense/inflammation genes by preterm infant microbiota in a humanized gnotobiotic mouse model. *PLoS One* **10**, e0124504 (2015).
39. J. Lu *et al.*, Effects of Intestinal Microbiota on Brain Development in Humanized Gnotobiotic Mice. *Sci Rep* **8**, 5443 (2018).
40. J. Lu *et al.*, Early preterm infant microbiome impacts adult learning. *Sci Rep* **12**, 3310 (2022).
41. S. J. Park, D. S. Im, Blockage of sphingosine-1-phosphate receptor 2 attenuates allergic asthma in mice. *Br J Pharmacol* **176**, 938-949 (2019).
42. M. J. Patil, S. Meeker, D. Bautista, X. Dong, B. J. Undem, Sphingosine-1-phosphate activates mouse vagal airway afferent C-fibres via S1PR3 receptors. *J Physiol* **597**, 2007-2019 (2019).
43. B. Thebaud *et al.*, Bronchopulmonary dysplasia. *Nat Rev Dis Primers* **5**, 78 (2019).
44. N. C. Stevens *et al.*, Alteration of glycosphingolipid metabolism by ozone is associated with exacerbation of allergic asthma characteristics in mice. *Toxicol Sci* **191**, 79-89 (2023).
45. R. Buda *et al.*, Dynamics of passive response to a sudden decrease in external osmolarity. *P Natl Acad Sci USA* **113**, E5838-E5846 (2016).
46. S. H. Mirhoseini, M. Nikaeen, H. Khanahmad, M. Hatamzadeh, A. Hassanzadeh, Monitoring of airborne bacteria and aerosols in different wards of hospitals - Particle counting usefulness in investigation of airborne bacteria. *Annals of Agricultural and Environmental Medicine* **22**, 670-673 (2015).
47. D. U. Park, J. K. Yeom, W. J. Lee, K. M. Lee, Assessment of the Levels of Airborne Bacteria, Gram-Negative Bacteria, and Fungi in Hospital Lobbies. *Int J Env Res Pub He* **10**, 541-555 (2013).
48. F. Schulte, J. Lingott, U. Panne, J. Kneipp, Chemical Characterization and Classification of Pollen. *Analytical Chemistry* **80**, 9551-9556 (2008).

49. A. Guedes, H. Ribeiro, M. Fernández-González, M. J. Aira, I. Abreu, Pollen Raman spectra database: Application to the identification of airborne pollen. *Talanta* **119**, 473-478 (2014).
50. D. J. Wang, J. C. White, Benefit of nano-enabled agrochemicals. *Nature Food* **3**, 983-984 (2022).
51. C. S. Ho *et al.*, Rapid identification of pathogenic bacteria using Raman spectroscopy and deep learning. *Nature Communications* **10**, (2019).
52. C. Wang *et al.*, Biomimetic olfactory chips based on large-scale monolithically integrated nanotube sensor arrays. *Nature Electronics*, (2024).
53. H. Weisbecker *et al.*, AI-Assisted Multimodal Breath Sensing System with Semiconductive Polymers for Accurate Monitoring of Ammonia Biomarkers. *Advanced Materials Technologies* **9**, (2024).
54. J. Li *et al.*, Aerodynamics-assisted, efficient and scalable kirigami fog collectors. *Nature Communications* **12**, (2021).
55. T. Li *et al.*, Scalable and efficient solar-driven atmospheric water harvesting enabled by bidirectionally aligned and hierarchically structured nanocomposites. *Nature Water* **1**, 971-981 (2023).
56. S. H. Nah *et al.*, Moisture Absorbing and Water Self-Releasing from Hybrid Hydrogel Desiccants. *Advanced Functional Materials* **34**, (2024).
57. T. S. Wong *et al.*, Bioinspired self-repairing slippery surfaces with pressure-stable omniphobicity. *Nature* **477**, 443-447 (2011).
58. G. Graeber *et al.*, Extreme Water Uptake of Hygroscopic Hydrogels through Maximized Swelling-Induced Salt Loading. *Advanced Materials* **36**, (2024).
59. Y. Zhong *et al.*, Bridging materials innovations to sorption-based atmospheric water harvesting devices. *Nature Reviews Materials*, (2024).
60. D. H. Ho, Y. Y. Choi, S. B. Jo, J. M. Myoung, J. H. Cho, Sensing with MXenes: Progress and Prospects. *Advanced Materials* **33**, (2021).
61. L. B. Huang *et al.*, Ultrasensitive, Fast-Responsive, Directional Airflow Sensing by Bioinspired Suspended Graphene Fibers. *Nano Letters* **23**, 597-605 (2023).
62. K. N. Wang *et al.*, A Polarity-Sensitive Ratiometric Fluorescence Probe for Monitoring Changes in Lipid Droplets and Nucleus during Ferroptosis. *Angewandte Chemie-International Edition* **60**, 15095-15100 (2021).
63. T. J. Dai *et al.*, An AIEgen as an Intrinsic Antibacterial Agent for Light-Up Detection and Inactivation of Intracellular Gram-Positive Bacteria. *Advanced Healthcare Materials* **10**, (2021).
64. J. J. Zhang *et al.*, Ultrasensitive point-of-care biochemical sensor based on metal-AIEgen frameworks. *Science Advances* **8**, (2022).